\documentclass[aps,pre,twocolumn,superscriptaddress,showkeys,showpacs]{revtex4-1}
\usepackage{blindtext}
\usepackage{tikz}
\usetikzlibrary{quantikz2}
\usepackage{algorithm}
\usepackage{algpseudocode}
\usepackage{amsmath}
\usepackage{amssymb}
\usepackage{nicefrac}
\usepackage{hyperref}
\usepackage{dsfont}
\usepackage{float}
\usepackage{graphicx}
\usepackage{tabularx}
\usepackage{braket}
\usepackage{subcaption}
\hypersetup{colorlinks = true, citecolor=blue}
\usepackage{comment}
\usepackage{soul}
\usepackage{bm}
\usepackage{amssymb}

\newcolumntype{P}[1]{>{\centering\arraybackslash}p{#1}}

\begin{document}
\title{Classically Augmented Zero-Noise Extrapolation}
\author{Timon Scheiber}
\email{timon.florian.scheiber@igd.fraunhofer.de}
\affiliation{Fraunhofer Institute for Computer Graphics Research IGD} 
\affiliation{Technical University of Darmstadt, 
Interactive Graphics Systems Group}

\begin{abstract}
We investigate a hybrid quantum-classical approach to quantum error mitigation. We propose Classically Augmented Zero-Noise Extrapolation, a hybrid error-mitigation method in which high-noise Richardson extrapolation nodes are replaced by classically simulated estimates. These classical nodes have negligible sampling variance but introduce deterministic simulation bias. We derive the resulting variance reduction under optimal shot allocation and show that, for linear node spacings and fixed index cutoff, the coefficient-level reduction can be exponential. We validate the prediction numerically using Pauli-propagation simulations and demonstrate a reduction in mean-squared error when the truncation bias is sufficiently small.
\end{abstract}

\maketitle
\section{Introduction}
Quantum computing is a potentially transformative technology for fields such as materials science and quantum chemistry~\cite{bauer2020quantum, babbush2026grand, santagati2024drug}. Although current-generation quantum computers can already perform certain calculations at the limit of what is tractable by classical methods~\cite{bluvstein2024logical, morvan2024phase, mazurenko2017cold, kim2023evidence}, 
some of these advantage claims have recently been challenged, as the underlying computations turned out to be reproducible by classical methods such as tensor networks~\cite{tindall2024efficient} or Pauli propagation~\cite{rudolph2023classical}.\\
Even though these devices are scientifically interesting, their usefulness for practical applications remains limited, the primary bottleneck being high error rates.
While the first demonstrations of error-corrected quantum computers are now emerging~\cite{bluvstein2024logical, yamamoto2026quantum}, the error rates of these early devices are still orders of magnitude above the levels required for most fault-tolerant algorithms.\\
An intermediate solution to the problem of errors in computation is the use of quantum error mitigation (QEM) techniques~\cite{cai2023quantum}. In general, QEM is an umbrella term for a class of algorithmic methods that reduce the bias of an (expectation value) estimation at the cost of increased sampling variance. Existing QEM approaches include Probabilistic Error Cancellation~\cite{temme2017error, van2023probabilistic}, Clifford Data Regression~\cite{czarnik2021error}, and Virtual Distillation~\cite{huggins2021virtual, koczor2021exponential}, among others~\cite{endo2018Practical}.

One of the most widely used QEM techniques is zero-noise extrapolation (ZNE)~\cite{temme2017error, kandala2019error, li2017efficient, giurgica2020digital, majumdar2023best, he2020zero}. ZNE mitigates errors by evaluating a circuit at several amplified noise levels and then extrapolating the resulting estimates to the zero-noise limit. In the small-error regime, this extrapolation can be performed efficiently using methods such as Richardson extrapolation. In Richardson extrapolation, the bias of the estimator can be reduced by increasing the number of extrapolation nodes~\cite{mohammadipour2025direct}. However, the variance of the extrapolated estimate also grows with the number of nodes.
A major bottleneck of Richardson-based error mitigation is the large sampling overhead induced by this variance amplification.
\\
Recently, several distinct error mitigation methods have been proposed that use classical pre- or post-processing to shift part of the computational overhead from the quantum computer to a classical computer~\cite{scheiber2025reduced, filippov2023scalable, majumder2026quantum, eddins2026computing}. In the same spirit, this work demonstrates that combining quantum and classical computation can lead to improved error mitigation.

We study the use of Pauli-propagation-based classical simulation to provide classically computed anchor points for Richardson extrapolation at high noise levels. In the proposed scheme, these high-noise nodes are replaced by classical estimates with negligible sampling variance but potentially nonzero approximation bias. Under optimal shot allocation, this leads to a reduction in the sampling variance of the extrapolated estimator. For linearly spaced nodes and a fixed cutoff, the resulting coefficient-level variance reduction can scale exponentially with the Richardson order. For Chebyshev-type spacings, the reduction is more modest but can still be practically relevant. In our numerical experiments, the reduction in sampling overhead can reach several orders of magnitude in favorable regimes.\\
We also use this setting to highlight a broader conceptual question. Which parts of the mitigation task genuinely require the quantum device, and which can be shifted to classical post-processing? We return to this point in Sec.~\ref{sec:Discussion}.\\

The remainder of this work is structured as follows. In Sec.~\ref{sec:ZNE} we review ZNE based on Richardson extrapolation. In Sec.~\ref{sec:CA-ZNE} we formally introduce Classically-Augmented ZNE, derive a closed-form expression for the variance reduction, and discuss several caveats associated with the classical simulation. Sec.~\ref{sec:Results} presents numerical simulations for two distinct Hamiltonians and examines the effect of the node spacing as well as the simulation bias. Finally, in Secs.~\ref{sec:Discussion} and~\ref{sec:CO} we discuss the implications of the method on what role the quantum computer plays in error mitigation and provide an outlook on future work.

\subsection{Main Contributions}
In this work, we introduce and analyze Richardson extrapolation in which
selected noisy expectation values are replaced by classically computed
nodes with zero sampling variance. We derive a closed-form expression for
the resulting reduction in sampling overhead and evaluate it for several
node-spacing schemes commonly used in zero-noise extrapolation. Our
analysis shows that incorporating zero-variance nodes can substantially
alter the variance properties of these schemes and suggests that node
spacings optimized for conventional zero-noise extrapolation need not
remain optimal in the classically augmented setting.
Finally, we briefly comment on the role the quantum computer plays in error mitigation if a sufficient set of noisy nodes can be simulated classically.
\section{zero-noise extrapolation}
\label{sec:ZNE}
Zero-noise extrapolation (ZNE) is a widely used technique for mitigating errors in noisy quantum computers~\cite{temme2017error}. The goal of ZNE is to estimate the ideal, noise-free expectation value
\begin{equation}
\langle O \rangle_{0} = \mathrm{Tr}\!\left(O\,\mathcal{C}_{\lambda=0}(\rho_0)\right) = \mathrm{Tr}\!\left(O\,\rho_{\lambda=0}\right),
\end{equation}
where $O$ is an observable and $\mathcal{C}_\lambda$ denotes a noisy quantum circuit with noise strength $\lambda$. To this end, the noise strength is systematically increased to obtain expectation values at several noise levels, which are then extrapolated to the zero-noise limit $\lambda \to 0$. In the low-noise regime, this extrapolation can be carried out efficiently using Richardson extrapolation.
The idea behind Richardson extrapolation for ZNE is to sample a set of $n$ distinct noise levels $\{\lambda_0, ...,\lambda_{n-1} \}$, often referred to as nodes, estimate the corresponding expectation values $\{ \text{Tr}(O\rho_{\lambda_0}), ...,\text{Tr}(O\rho_{\lambda_{n-1}}) \}$, and then determine the unique polynomial $f(\lambda)$ of degree $n - 1$ that interpolates the resulting data. Throughout this work, $n$ denotes the number of Richardson nodes.
The interpolating polynomial has degree $n-1$, which we refer to as
the extrapolation order.
The resulting interpolating polynomial can then be written as
\begin{equation}
\label{eq:rich}
    f(\lambda) = \sum_{i=0}^{n-1} \text{Tr}(O\rho_{\lambda_i}) \,\gamma_i(\lambda)
\end{equation}
where the functions $\gamma_i(\lambda)$ are the Lagrange basis polynomials,
\begin{equation}
    \gamma_i(\lambda) = \prod_{k \neq i} \frac{\lambda - \lambda_k}{\lambda_i - \lambda_k}.
\end{equation}
Here, the amplified noise levels are written as $\lambda_k=\lambda_0 x_k$, where $\lambda_0$ is the native hardware noise level and the factors $x_k$ define the chosen node spacing.
By evaluating the interpolating polynomial $f$ at the zero-noise limit, one obtains an approximation to the ideal expectation value
\begin{equation}
   \langle O \rangle_0 \approx f(0) = \sum_{i=0}^{n-1}  \text{Tr}(O\rho_{\lambda_i}) \prod_{k \neq i} \frac{\lambda_k}{\lambda_k -\lambda_i}.
\end{equation}
To perform ZNE, one generally needs to estimate the expectation values corresponding to the noise levels $\lambda_i$ on a quantum computer by measuring the observable of interest $N$ times. According to the central limit theorem, the resulting sample mean approximates the true expectation value, with variance
\begin{equation}
    \text{Var}[\hat{O}] = \frac{\sigma_O^2}{N} \quad \text{with} \quad \sigma_O = \sqrt{\langle O^2 \rangle - \langle O \rangle^2}
\end{equation}
where $\hat{O}$ is an estimator for the expectation value of $O$.

Let $\hat{O}_{\lambda_i}$ denote the empirical estimator of $\langle O\rangle_{\lambda_i}$ obtained from $N_i$ shots. Assuming independent estimates across noise levels, the variance of the extrapolated estimator
\begin{equation}
    \hat{f}(0)=\sum_{i=0}^{n-1}\gamma_i(0)\,\hat{O}_{\lambda_i}
\end{equation}
is
\begin{equation}
\label{eq:var}
    \mathrm{Var}[\hat{f}(0)]
    =
    \sum_{i=0}^{n-1}\gamma_i(0)^2\,\mathrm{Var}[\hat{O}_{\lambda_i}]
    =
    \sum_{i=0}^{n-1}\frac{\sigma_{O,i}^2}{N_i}\gamma_i^2,
\end{equation}
where we use the shorthand notation $\gamma_i=\gamma_i(0)$.
As noted in Ref.~\cite{krebsbach2022optimization}, the variances $\sigma_{O,i}^2$ are often of comparable magnitude across nodes. For analytical simplicity, we therefore adopt the approximation $\sigma_{O,i}=\sigma_O$ when deriving the closed-form reduction formulas below. We explicitly note when this simplifying assumption is used.
Eq.~\eqref{eq:var} shows that the extrapolated variance is controlled by the magnitudes of the Lagrange weights $\gamma_i$, which in turn depend strongly on the chosen node spacing. This variance boost is commonly expressed in terms of the sampling overhead~\cite{cai2023quantum, krebsbach2022optimization, giurgica2020digital}
\begin{equation}
    \Lambda^2 = \left(\sum_{i=0}^{n-1} |\gamma_i| \right)^2 = \left(\sum_{i=0}^{n-1} \left|\prod_{k \neq i}\frac{\lambda_k}  {\lambda_k -\lambda_i} \right| \right)^2.
\end{equation}
The sampling overhead describes the scaling factor that is needed to compensate for the variance boost induced by the Richardson extrapolation.
\\
The large sampling overhead of the estimator is one of the main factors determining the usefulness of the method.
One way to directly reduce this sampling overhead is through importance sampling. For a fixed total shot budget $N_{\text{shots}}$, it is optimal to allocate more shots to nodes that are expected to contribute more strongly to the total variance, i.e.,~nodes that have a large interpolation weight $N_i \propto|\gamma_i|$.
Under this assumption, a straightforward calculation shows that the optimal shot allocation is given by~\cite{cai2021practical, krebsbach2022optimization}
\begin{equation}
    N_i = N_\text{shots}\frac{|\gamma_i|}{\sum_{k=0}^{n-1} |\gamma_k|}
\end{equation}
which is obtained by directly minimizing the variance.
Since the node variances are generally not known before sampling this expression relies on the assumption $\sigma_{O, i} \approx \sigma_O$.
Hence, by allocating more shots to nodes that contribute more strongly to the total variance, the overall variance can be reduced. A more detailed analysis of optimal node spacing and importance sampling is given in Refs.~\cite{krebsbach2022optimization, giurgica2020digital, mohammadipour2025direct}.

\section{Classically augmented zero-noise extrapolation}
\label{sec:CA-ZNE}
\begin{figure}
    \centering
    \includegraphics[width=\linewidth]{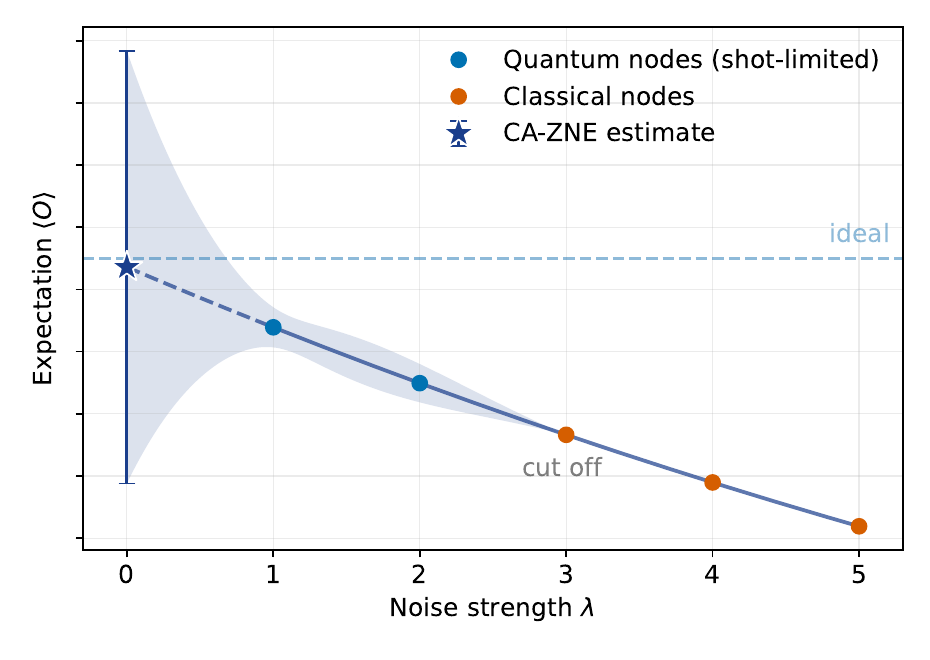}
    \caption{Schematic of CA-ZNE. Standard ZNE estimates the noise-free expectation value by evaluating a noisy quantum circuit at several amplified noise levels and extrapolating the resulting expectation values to the zero-noise limit. Because Richardson extrapolation amplifies statistical fluctuations, the resulting estimator can have large sampling variance and a potential extrapolation bias. In CA-ZNE, nodes above a cutoff noise level are replaced by classically simulated estimates, which act as zero-sampling-variance anchor points for the extrapolation. This can stabilize the estimator and reduce variance, at the cost of a possible deterministic simulation bias.}
    \label{fig:CA_ZNE}
\end{figure}
In general, Richardson extrapolation suffers from increasing variance as the number of nodes grows. This increase can be mitigated by shifting part of the computational overhead from the quantum computer to a classical computer. In particular, some of the nodes in Eq.~\eqref{eq:rich} can be replaced by classical estimates,
\begin{equation}
    \hat{f}_\text{aug}(0) = \sum_{i=0}^{c-1}\hat{O}_{i}^\text{quant} {\gamma_i} + \sum_{i=c}^{n-1}  \hat{O}_{i}^\text{class}{\gamma_i}
\end{equation}
where the first $c$ nodes are estimated on the quantum device and the remaining nodes are supplied by a potentially biased classical simulator with negligible variance $\text{Var}[\hat{O}_{i}^\text{class}] \approx0$. We refer to this hybrid method as Classically Augmented Zero-Noise Extrapolation (CA-ZNE). We provide a pictorial representation of the main idea in Fig.~\ref{fig:CA_ZNE}.\\
This replacement is particularly attractive at high noise levels, where noisy circuits can become easier to simulate classically, for example via truncated Pauli propagation~\cite{angrisani2026simulating, schuster2025polynomial}, albeit with a controlled approximation bias.\\
The main advantage of classically simulating these nodes, rather than evaluating them on a quantum computer, is that for a deterministic algorithm they have essentially zero sampling variance.
If the classical backend is deterministic, or if its internal numerical error is negligible compared with shot noise, the sampling variance of the hybrid estimator reduces to
\begin{equation}
    \mathrm{Var}[\hat{f}_{\mathrm{aug}}(0)]
    =
    \sum_{i=0}^{c-1}\frac{\sigma_{O,i}^2}{N_i}\gamma_i^2 
    \leq 
    \sum_{i=0}^{n-1}\frac{\sigma_{O,i}^2}{N_i}\gamma_i^2 
\end{equation}
which marks a direct variance improvement over the fully quantum estimator. It should however be noted that these nodes might be biased due to imperfect simulation.\\

As an additional benefit, replacing the data points beyond the cutoff by classical estimates frees a shot budget of $N_\text{add} = \sum_{i=c}^{n-1} N_i$. By reallocating this freed shot budget to the nodes sampled on the quantum device, the variance can be reduced even further.
The resulting variance ratio $R$ can then be expressed as
\begin{equation}
\label{eq:red}
    R = \frac{\sum_{j=0}^{c-1} \nicefrac{\gamma_j^2 \sigma_{O,j}^2}{\tilde{N}_j} }{\sum_{i=0}^{n-1} \nicefrac{\gamma_i^2\sigma_{O,i}^2}{N_i} }
\end{equation}
where $\tilde{N}_i =  N_\text{shots}\frac{|\gamma_i|\sigma_{O,i}}{\sum_{k=0}^{c-1} |\gamma_k|\sigma_{O,k}}$ is the theoretically optimal shot allocation obtained by distributing the full budget over only the first $c$ data points. If optimal shot allocation is used for both approaches at the same total shot budget,
Eq.~\eqref{eq:red} simplifies to
\begin{equation}
\label{Eq.:red}
    R = \left(\frac{\sum_{j=0}^{c-1} |\gamma_j|\sigma_{O,j}}{\sum_{i=0}^{n-1} |\gamma_i|\sigma_{O,i}}\right)^2
\end{equation}
which is independent of the shot budget $N_\text{shots}$. In practice the exact values of the $\sigma_{O,i}$ cannot be known a priori and the theoretically optimal shot allocation cannot be performed. Under the assumption of equal $\sigma_{O,i}$ the optimal shot allocation is given by $\tilde{N}_i =  N_\text{shots}\frac{|\gamma_i|}{\sum_{k=0}^{c-1} |\gamma_k|}$ and leads to the reduction
\begin{equation}
\label{Eq.:red}
    R = \left(\frac{\sum_{j=0}^{c-1} |\gamma_j|}{\sum_{i=0}^{n-1} |\gamma_i|}\right)^2
\end{equation}
which depends only on the interpolation weights.\\
It is noteworthy that, at the level of Richardson coefficients, the variance-reduction mechanism depends only on the chosen node spacing and cutoff and is thus circuit independent. Its practical usefulness, however, remains circuit- and hardware-dependent through the attainable classical bias and the location of a suitable cutoff.\\

The classical nodes reduce variance but introduce a deterministic approximation bias. We show in Appendix~\ref{Appendix:Bias} that the bias in the Richardson extrapolation decomposes into the polynomial extrapolation bias as well as the classical node bias
\begin{equation}
    \mathbb{E}[\hat{f}_\text{aug}(0)] - \langle O\rangle_0 = \Delta_\text{extrap} + \sum_{i=c}^{n-1} \delta_i \gamma_i
\end{equation}
where $\Delta_\text{extrap}$ is the extrapolation bias (cf. Appendix~\ref{Appendix:Bias} ) and the $\delta_i$ are the per node simulation biases. If these simulation biases are too large, the reduction in variance may not compensate for the additional bias.\\

In order to see when CA-ZNE is advantageous over direct ZNE it is relevant to consider the bias–variance tradeoff. One quantity that captures this trade off is the mean-squared error $\text{MSE}= \mathbb{E}[(\hat{y} -y)^2] $, 
which decomposes into 
\begin{equation}
    \text{MSE}(\hat{y}) = \text{Bias}(\hat{y})^2 + \text{Var}(\hat{y}).
\end{equation}
The MSE of ZNE is then explicitly given by the relation
\begin{equation}
    \text{MSE}_\text{ZNE} = \Delta_\text{extrap}^2 + \sum_{i=0}^{n-1} \frac{\sigma_O^2}{N_i}{\gamma_i}^2.
\end{equation}
In contrast for CA-ZNE the MSE is given by
\begin{equation}
    \text{MSE}_\text{CA-ZNE} = \left(\Delta_\text{extrap} + \sum_{i=c}^{n-1} \delta_i \gamma_i \right)^2 + \sum_{i=0}^{c-1} \frac{\sigma_O^2}{\tilde{N}_i}{\gamma_i}^2.
\end{equation}
Therefore, CA-ZNE improves upon standard ZNE in mean-squared error whenever the additional classical bias is sufficiently small relative to the variance reduction obtained from replacing and reallocating the high-noise nodes.

\subsection{Noise model}
\label{NoiseModel}
In the last paragraph we have considered CA-ZNE at a high level.
At a lower level, for the classically calculated data points to integrate with the quantum data points, it is important to use the same noise model as the quantum computer in the classical simulation.

We perform this hybrid zero-noise extrapolation by proposing the use of a \textit{digital twin} that runs using a learned noise model from the quantum computer.

In this work we consider a sparse Pauli Lindblad (SPL) noise model~\cite{van2023probabilistic} which is a reasonable approximation for noise on quantum devices with limited connectivity. The SPL model is generated by a Lindbladian of the form
\begin{equation}
\label{eq:Lindblad}
    \mathcal{L}(\rho) =\sum_{k\in \mathcal{K}} \lambda_k (P_k\rho P_k - \rho).
\end{equation}
where the set $\mathcal{K}$ runs over a small subset of Pauli jump operators $P_k$ and generators $\lambda_k$ that closely follow the hardware connectivity graph.
The noise channel generated by this Lindbladian is then given by 
\begin{equation}
     \mathcal{N} (\rho) = \prod_{k \in \mathcal{K}}(w_k \mathcal{I}(\cdot) + (1-w_k) \mathcal{P}_k(\cdot)) \rho
\label{eq:SPL_noise}
\end{equation}
with $\mathcal{P}_k(\cdot)=P_k(\cdot)P_k$ being a short hand notation.
This model is attractive because it can be learned efficiently~\cite{malekakhlagh2025efficient, belkin2026better} and is already used in noise-aware mitigation workflows~\cite{van2023probabilistic, kim2023evidence, filippov2023scalable}.\\
We do not model errors arising from mismatch between the learned noise model and the true device noise. This omission does not affect the synthetic simulations studied here, but it is an important limitation in the experimental setting. In practice, this assumption is most natural when comparing against mitigation baselines, such as probabilistic error amplification, that already rely on a learned noise model.
CA-ZNE can be added naturally to workflows that already rely on a learned noise model, with no additional quantum sampling cost beyond the remaining quantum nodes and with the main extra cost being the classical simulation.\\
A potential noise learning procedure including an error analysis is provided by Ref.~\cite{van2023probabilistic}.

\subsection{Caveats}

It is important to note that the classical nodes are computed only to finite precision and therefore introduce bias. A natural question is whether this bias is acceptable in practice. We address this by defining a crossover point at which the sampling error of a quantum node exceeds the truncation error of the corresponding classical simulation.

\begin{figure}[H]
    \centering
    \includegraphics[width=\linewidth]{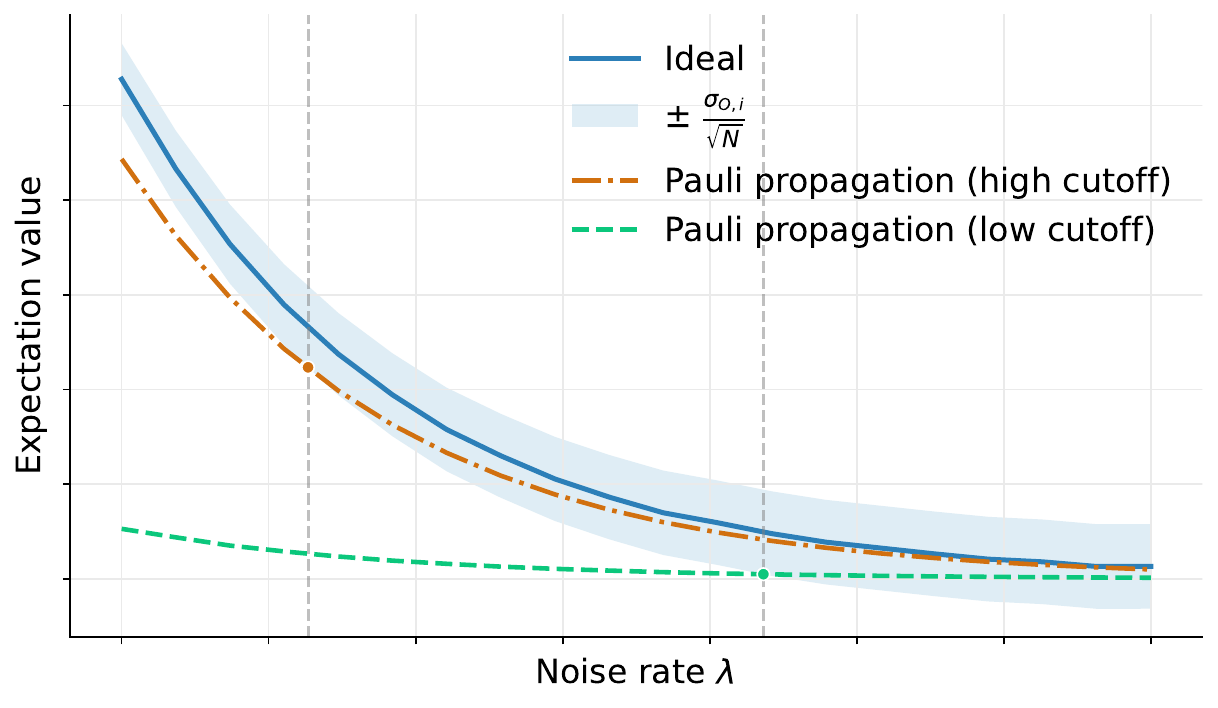}
    \caption{Illustration of the crossover condition between shot noise and classical truncation error. The shaded regions indicate the sampling uncertainty of the quantum estimates at fixed shot budget. At the points where the Pauli propagation curves intersect the shaded region the simulation bias becomes smaller than the expected sampling error. The higher the cutoff in the Pauli propagation the more exact the simulation will become and the intersection occurs earlier.}
    \label{fig:Crossover}
\end{figure}

In general, an estimate of an expectation value obtained from $N$ samples has statistical error
\begin{equation}
    \Delta_\text{sample} = \frac{\sigma_O}{\sqrt{N}}.
\end{equation}
By contrast, the classical simulation is associated with a truncation error $\Delta_\text{trunc}$. We therefore identify the crossover point as the regime in which the truncation error becomes smaller than the sampling error, as illustrated in Fig.~\ref{fig:Crossover}
\begin{equation}
\label{eq:cross}
    \Delta_\text{trunc} \leq \Delta_\text{sample}.
\end{equation}

In practice, neither the true truncation bias nor the exact shot noise are known a priori, so Eq.~\eqref{eq:cross} should not be interpreted as a directly computable cutoff rule. Instead, it describes the regime in which replacing a quantum node by a classical node is beneficial. Operationally the bias of the nodes is controlled by the amount of classical resources one is able to allocate. Given a classical wall-clock-time budget the cutoff could in principle be estimated using a convergence analysis.\\

Beyond this crossover point, it becomes more efficient to simulate the corresponding nodes classically. For Pauli propagation with layer-wise weight truncation under a local depolarizing noise model, there exists a polynomial-time simulation algorithm~\cite{schuster2025polynomial} together with an explicit upper bound on the approximation error
\begin{equation}
\label{truncation_error}
\Delta_\text{trunc}
= \left|\langle O \rangle_\text{class} - \langle O \rangle_\text{exact}\right|
\leq e^{-\nu(l+1)} \sqrt{d+1}\,\|O\|_F .
\end{equation}
Here, $\langle O \rangle_\text{class}$ denotes the estimate returned by the algorithm, $\nu$ is the noise rate per qubit per layer, $l$ is the truncation weight threshold, $d$ is the circuit depth, and $\|O\|_F$ is the Frobenius norm of the observable $O$.
It is important to note that no general bound is known that applies to all quantum circuits, including worst-case instances. The bounds derived in Ref.~\cite{schuster2025polynomial} rely on the assumption that the error is bounded in root-mean-square error over an ensemble $E$ of input states $\{\rho_\text{in}\}$. Consequently, these bounds hold only on average over computational-basis input states. Similar results can also be obtained under the assumption of uncorrelated gate parameters~\cite{angrisani2026simulating}.

In Appendix~\ref{Appendix:NoiseModel}, we show that the damping of Pauli terms under the SPL noise model can be upper-bounded by a layer-wise depolarizing channel with effective noise strength $\nu_\text{eff} = 4\lambda_\text{min}$, as assumed in Ref.~\cite{schuster2025polynomial}. Here, $\lambda_\text{min}$ denotes the smallest measured single-qubit rate of the SPL Lindbladian $\mathcal{L}$.

Under this assumption we are able to apply the algorithm of Ref.~\cite{schuster2025polynomial} with the effective noise rate $\nu_\text{eff}$ to obtain a classical estimate together with a conservative bias bound.
Since the simulation error in Eq.~\eqref{truncation_error} decreases exponentially with $\nu$, the approximation becomes rapidly more accurate at higher noise levels.

The bounds presented here are generally rather loose. In Appendix~\ref{Appendix:NoiseModel}, we approximate the effect of the SPL noise model by an effective layer-wise depolarizing channel. Since this approximation is based on the minimum single-qubit rate, the resulting bound is conservative. The actual SPL noise typically induces stronger damping than the depolarizing channel used to derive the bound.

\subsection{Bounds on the bias}
\label{sec:biasbound}
To use the proposed method responsibly in practice, one must control the additional bias introduced by the classically simulated nodes. Under the assumptions of Ref.~\cite{schuster2025polynomial} and with effective noise rate $\nu_\text{eff}$, we obtain the following upper bound on the expected bias for some $\xi\in[0,x_{n-1}]$
\begin{equation}
    ~\begin{split}
    \left|\text{Bias}[\hat{f}_\text{aug}] \right|  &\leq
    \left|\frac{ g^{(n)}(\xi)}{(n)!}\prod_{j=0}^{n-1} x_j \right| \\
    &+\sqrt{d+1} ||O||_F\sum_{i=c}^{n-1}  e^{-\nu_\text{eff}x_i(l+1)}~|  \gamma_i|
\end{split}
\end{equation}
which decomposes into the polynomial interpolation error and the classical simulation error. Here, $g(x)=\langle O\rangle_{\lambda_0 x}$ denotes the exact noisy expectation-value curve being approximated by the polynomial $f$.
These guarantees are root-mean-square statements over the input ensembles considered in Ref.~\cite{schuster2025polynomial} and are not worst-case per-instance guarantees for arbitrary circuits and inputs. Per-instance 
guarantees for fixed inputs are fundamentally 
unattainable for general circuits~\cite{schuster2025polynomial, 
shor1996fault, aharonov1997fault}, since for a fixed input and 
sufficiently low noise rates QEC can be performed in principle. In contrast, 
in the high-noise regime, expectation values are exponentially 
dominated by low-weight Pauli operators~\cite{schuster2025polynomial, 
aharonov1996polynomial}, enabling efficient classical simulation.

We provide the exact derivation of the bias bound as a short proof in Appendix~\ref{Appendix:Bias}.

\subsection{Classical simulation via Pauli propagation}
To classically simulate the high-noise ZNE nodes, it is essential to use a method that can faithfully represent a learned hardware noise model while also allowing the noise strength to be increased in a controlled manner. A method particularly well suited to this task is Pauli propagation~\cite{schuster2025polynomial,angrisani2026simulating,pauli-prop}. Pauli propagation aims to obtain a classical estimate of the expectation value $\mathrm{Tr}\!\left(O\,C(\rho)\right)$ by means of \textit{operator backpropagation}. In this approach, the observable is propagated backward through the adjoint circuit in the Heisenberg picture, i.e., one considers $C^\dagger(O)$. To do so, the observable is first expanded in the Pauli basis
\begin{equation}
    O = \sum_{i=1}^m c_i P_i,
\end{equation}
where $P_i$ are Pauli strings. It is important to note that Pauli propagation assumes that the observable has an efficient Pauli decomposition, which does not include exponentially many terms. Each of the $m$ Pauli terms is then propagated backward individually, which yields
\begin{equation}
    \tilde{O} = C^\dagger(O) = \sum_{i=1}^m c_i\, C^\dagger(P_i).
\end{equation}
Since the Pauli operators form a complete basis for the operator space, the propagated observable $C^\dagger(O)$ can again be expressed as a Pauli decomposition.

To implement this backpropagation efficiently, the circuit is written as a sequence of quantum operations,
\begin{equation}
     \mathcal C^\dagger(O) = \mathcal E_0^\dagger\circ\cdots\circ\mathcal E_L^\dagger(O).
\end{equation}
Here, the maps $\mathcal{E}_j$ are not necessarily unitary and may also represent noise processes.

For the unitary part of the circuit, it is often advantageous to express the evolution as a sequence of Clifford gates $\mathcal{C}$ and Pauli rotation gates $R_G$, which form a universal gate set. This decomposition is useful because a Clifford gate maps any Pauli operator to another Pauli operator,
\begin{equation}
    \mathcal{C}^\dagger(P) = P'.
\end{equation}
Clifford gates are therefore particularly favorable for simulation, since they conserve the number of Pauli terms in the expansion and thus lead to only constant memory overhead.

For non-Clifford gates, by contrast, there is no guarantee that the number of Pauli terms will remain constant. For Pauli rotation gates, the action is given by
\begin{equation}
    R_G^\dagger(P) = P \qquad \text{if} \qquad [P, G] = 0,
\end{equation}
and
\begin{equation}
    R_G^\dagger(P) = \cos(\theta) P + \sin(\theta) P' \qquad \text{if} \qquad [P, G] \neq 0.
\end{equation}
Thus, if $P$ commutes with the generator $G$ of the rotation gate, it remains unchanged. If $P$ and $G$ do not commute, the Pauli operator is mapped to a sum of Pauli terms, a phenomenon known as \textit{branching}. When too many such branching gates are present, the required memory grows exponentially. To avoid this exponential growth, Pauli terms whose coefficients $c_i$ fall below a chosen threshold are truncated, which leads to a more efficient but potentially biased solution.
\\
For Pauli-propagation algorithms, one possible runtime diagnostic of the bias is to accumulate the magnitude of discarded Pauli coefficients, which gives a conservative a posteriori upper bound on the simulation error~\cite{rudolph2025pauli}. This bound can be loose because it ignores cancellations.
Apart from this the bias can also be estimated empirically by performing a convergence analysis~\cite{gharibyan2025practical}.

We perform an empirical convergence analysis for the transverse-field
Ising model as an example in section~\ref{sec:Ising}. We do not, however, integrate this diagnostic
into an adaptive cutoff-selection protocol, which we leave to future work
\subsubsection{Interaction with SPL noise}

Pauli propagation is particularly well suited for simulating the SPL noise model because of its simple action on Pauli observables. The SPL channel is diagonal in the Pauli basis, so each Pauli operator $P$ is an eigenoperator of $ \mathcal{N}$ with eigenvalue (cf.~Eq.~\eqref{eq:Lindblad})
\begin{equation}
    \mathcal{N}(P) = e^{-2\sum_{k\in \mathcal{K}} \lambda_k \langle P, P_k\rangle}\, P,
\end{equation}
where $\langle P, P_k\rangle$ denotes the symplectic inner product, which equals $0$ if the two Pauli operators commute and $1$ if they anticommute.\\
The same effect extends naturally to the adjoint noise channel $\mathcal{N}^\dagger$ which is considered in most Pauli propagation implementations.

Thus, each Pauli coefficient is exponentially damped by the noise channel. Coefficients that are strongly suppressed can therefore be truncated with only a small effect on the expectation value, thereby improving computational efficiency.\\


\subsection{Node spacing}
\label{sec:node_spacing}
\begin{figure}
    \centering
    \includegraphics[width=\linewidth]{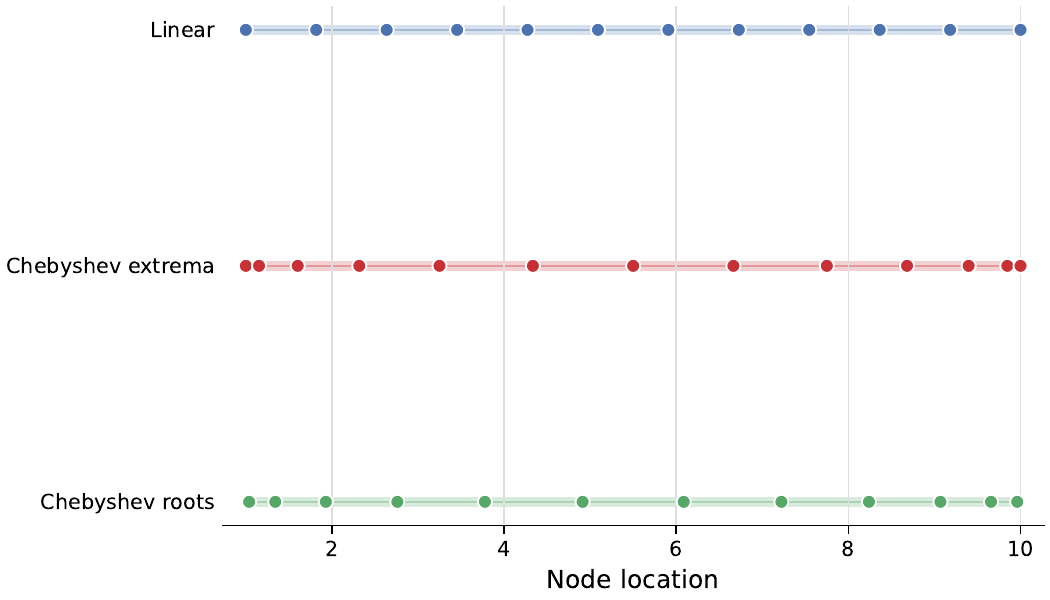}
    \caption{Graphical representation of the three node spacings considered in this work on the interpolation interval $[1,10]$. The spacings are a \textit{linear} node spacing and the two spacings closely aligned with the \textit{Chebyshev} nodes, namely the extrema and the roots of the polynomial of degree $n$. As highlighted by the figure the Chebyshev nodes cluster at the end points of the interpolation interval.}
    \label{fig:spacings}
\end{figure}
Apart from the hybrid simulation, ZNE relies on Richardson extrapolation.
One of the main limitations of Richardson extrapolation is the blow-up in variance, which is primarily determined by the spacing between the nodes $|x_i - x_k|$.
The most straightforward choice, used in many current applications such as the \textit{Mitiq} software package~\cite{mitiq}, is a linear node spacing
\begin{equation}
    x_j = 1 + h \cdot j
\end{equation}
with step size $h$.
In principle, such node sets are unbounded, but in many applications it is preferable to restrict them to a finite interval $[1,B]$ in order to avoid reaching the noise floor~\cite{koester2026benchmarking}. Even though they are potentially non-optimal, 
linear spacings remain attractive in practice because they can be implemented straightforwardly using standard noise-amplification techniques such as gate folding~\cite{he2020zero, giurgica2020digital} or pulse stretching~\cite{kim2023scalable}. Despite their practical simplicity and continued widespread use, linearly spaced nodes are known to be poorly conditioned for polynomial interpolation and may therefore also lead to unstable extrapolation behavior.

Recent work~\cite{krebsbach2022optimization, mohammadipour2025direct} focuses on the use of the Chebyshev node spacing which is better suited for polynomial interpolation. The term Chebyshev nodes is used rather ambiguously in the literature and might refer to the roots of the Chebyshev polynomials of the first kind $T_{n}(x)$ or their extrema.\\
We first consider the roots of the Chebyshev polynomials, which are given by
\begin{equation}
    \tilde{x}_j = \cos\left(\frac{(2j+1) \pi }{2n}\right), \quad j\in[n-1, n-2, ..., 0]
\end{equation}
and describe the zeros of the polynomial in the open interval (-1,1).
These nodes can then be mapped to the interval of interest via an affine transform
\begin{equation}
    \Phi(\tilde{x}_j) = \frac{B-1}{2}\tilde{x}_j+ \frac{B+1}{2},
\end{equation}
which gives rise to the node spacing $x_j = \Phi(\tilde{x}_j)$.\\

Another study has shown that, under certain conditions, the optimal node spacing is given by the so-called tilted Chebyshev nodes~\cite{hoel1964optimal, krebsbach2022optimization}
\begin{equation}
    x_j = 1 + \frac{\sin^2\left(\frac{j\pi}{2(n+1)}\right)}{\sin^2\left(\frac{\pi}{2(n+1)}\right)}(x_1 - 1)
\end{equation}
which minimize the bias for a given maximum sampling overhead under the constraint that $x_0 = 1$. In contrast to the first considered spacing, these nodes correspond to the extrema of the Chebyshev polynomials which are tilted, i.e.~n is increased by one and the last node is excluded~\cite{krebsbach2022optimization}. To ensure comparability of both  spacings in this work, we chose $x_1$ adaptively to restrict the full set of nodes to the interval $[1,B]$.
In contrast to the linear spacing, implementing these spacings typically requires a higher level of control than simple gate folding or pulse stretching, for example through probabilistic error amplification~\cite{li2017efficient, kim2023evidence} and might thus be especially well suited for the hybrid approach.

We give a graphical representation of the node spacings considered in this work in Fig.~\ref{fig:spacings}.\\

\section{Results}
\label{sec:Results}
\subsection{Simulation}
To validate our results we perform a numerical simulation of a quantum computer via the \textit{Qiskit Aer} simulator. We investigate the results on two different examples to highlight the utility to different NISQ algorithms.\\
For all simulations, we use the Pauli propagation algorithm of Ref.~\cite{pauli-prop}, which is conceptually similar to the method of Ref.~\cite{schuster2025polynomial} but employs a truncation parameter, \textit{max terms}, to keep the simulation overhead manageable. The Qiskit implementation controls the simulation cost through \textit{max terms}, which is not identical to the weight cutoff $l$ appearing in the theoretical bound. We therefore use the bound of section~\ref{sec:biasbound} only as motivation and interpret the numerical truncation study empirically.

\subsection{Simulation of the Transverse Field Ising model}
\label{sec:Ising}
In the initial tests, we consider the simulation of the transverse-field Ising model on 12 qubits for 4 Trotter steps, where each Trotter step is followed by a layer of local depolarizing channels.
The Hamiltonian of the transverse-field Ising model is given by
\begin{equation}
    H = J\sum_{(i, j) \in E} Z_iZ_j + h\sum_i X_i
\end{equation}
where $E$ denotes the set of edges defining the connectivity of the Hamiltonian, $J$ denotes the spin-spin coupling and $h$ is the external field strength. We select a $3 \times 4$ grid with a coupling strength $J = 1$ and external field of $h=0.7$ and a full evolution time of $t=1$. In our simulations, we study the expectation value of the two-qubit pair-correlation observable
\begin{equation}
    O = \sum_{(i, j) \in E} Z_iZ_j.
\end{equation}

\begin{figure}[h]
    \centering
    \includegraphics[width=\linewidth]{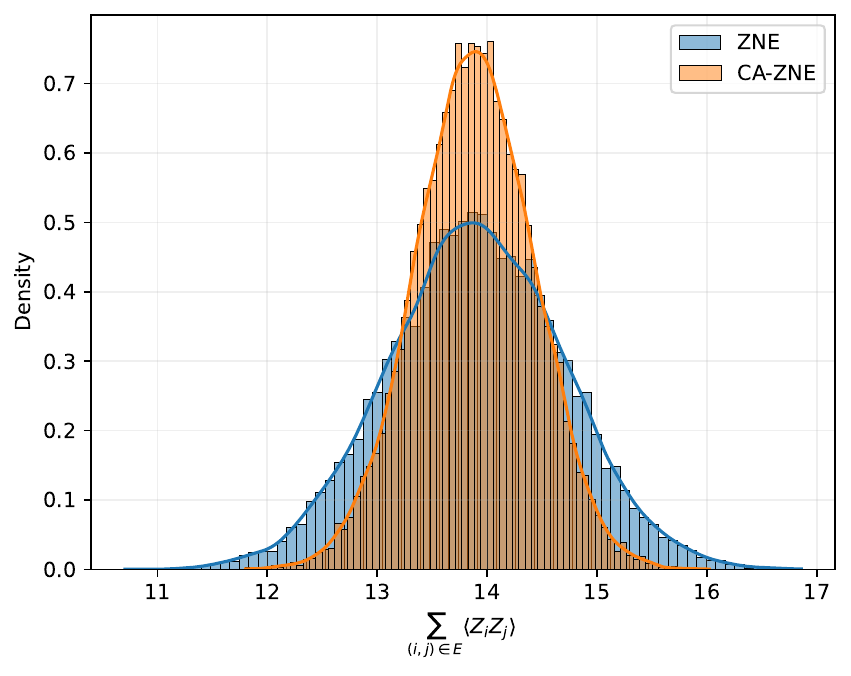}
    \caption{Obtained histogram by performing 15 000 runs of independent simulation for both methods. The histogram shows the distribution of the obtained expectation values from independent ZNE runs for CA-ZNE and ZNE. For ZNE we obtain a MSE of around $0.65$ whereas for CA-ZNE we obtain $0.3$. Our results show a bias of $0.129$ for ZNE and $0.141$ for CA-ZNE.}
    \label{fig:Histogram}
\end{figure}

\begin{figure*}[htp!]
    \centering
    \includegraphics[width=0.9\linewidth]{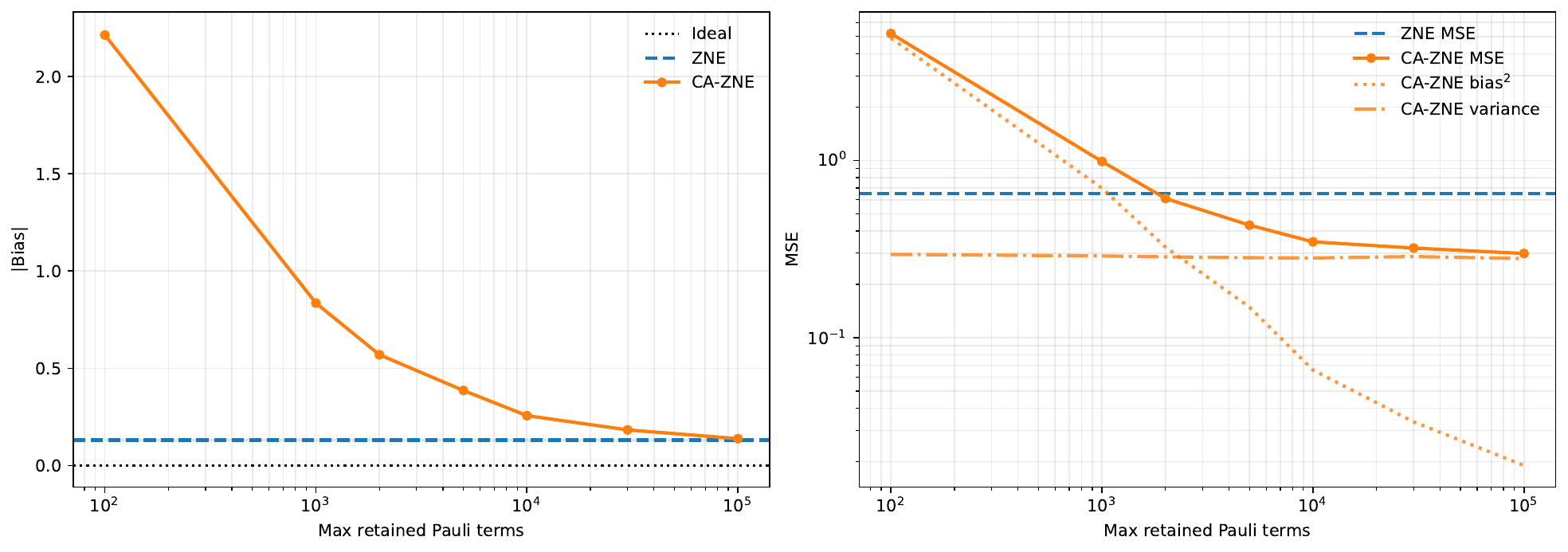}
    \caption{Convergence of CA-ZNE with the Pauli-truncation budget for the transverse-field Ising model. The x-axis denotes the maximum number of Pauli terms retained in the Pauli-propagation simulation. Left: empirical bias relative to the noiseless reference value. For small truncation budgets, CA-ZNE exhibits a large truncation-induced bias, which decreases rapidly as more Pauli terms are retained and approaches the fixed-order ZNE baseline. Shaded regions indicate the spread over repeated estimates. Right: corresponding mean-squared error (MSE). As the truncation bias is reduced, the CA-ZNE MSE drops and approaches the improvement set by the reduced estimator variance, eventually outperforming the pure ZNE baseline.}
    \label{fig:ConvergencePlot}
\end{figure*}

To verify the theoretical prediction we perform a Monte Carlo estimate of the variance reduction by performing 15 000 independent runs of ZNE with linearly scaled noise starting from $\lambda_0 = 0.001$ up to $\lambda_3 = 0.004$ for both direct ZNE as well as CA-ZNE with two quantum and two classical nodes. For both simulations a shot budget of $N_\text{shots} = 10^4$ was chosen. All simulations were performed on the density operator level in order to save computation time. The results are presented in Figure~\ref{fig:Histogram}. To investigate the limit of truncation error on the results, the simulation was performed for a maximum of $10^4$ distinct Pauli strings using the Qiskit Pauli propagation implementation~\cite{pauli-prop}, which is only a small fraction of the $4^{12} \approx 1.7 \cdot 10^7$ possible Pauli strings for the considered circuit. We investigate the effect of different truncation levels in Fig.~\ref{fig:ConvergencePlot}.

For the experiment we find that the classical augmentation and shot reallocation greatly reduced the variance of the estimate by $1 - R = 55.5 \%$ with only a slight increase in bias. Further, for the collected sample we find a mean square error
$\text{MSE}_\text{direct} = 0.65$ for direct ZNE and $\text{MSE}_\text{augmented} = 0.3$ for CA-ZNE.\\
Most of the difference in bias in our estimate can be eliminated by increasing the number of distinct Pauli strings in the propagation algorithm. In contrast the variance reduction has an analytically closed form (see Eq.~\eqref{Eq.:red}). Analytically we obtain a ratio of $R = 0.444$, which closely matches the variance ratio from our data $R = 0.445$. The small deviation follows most likely from the finite sample size or from small deviations of the sampling errors $\sigma_{i,O}$ for the different noise levels.
\\

\subsection{Simulation of the Schwinger model}
As an additional example, we consider the simulation of the dynamics of the Schwinger model, which is a widely used benchmark for quantum simulation in high-energy physics. The Schwinger model describes quantum electrodynamics on a lattice in one spatial and one temporal dimension and captures the coupling between electrons and photons.

\begin{figure}[htp!]
    \centering
    \includegraphics[width=1\linewidth]{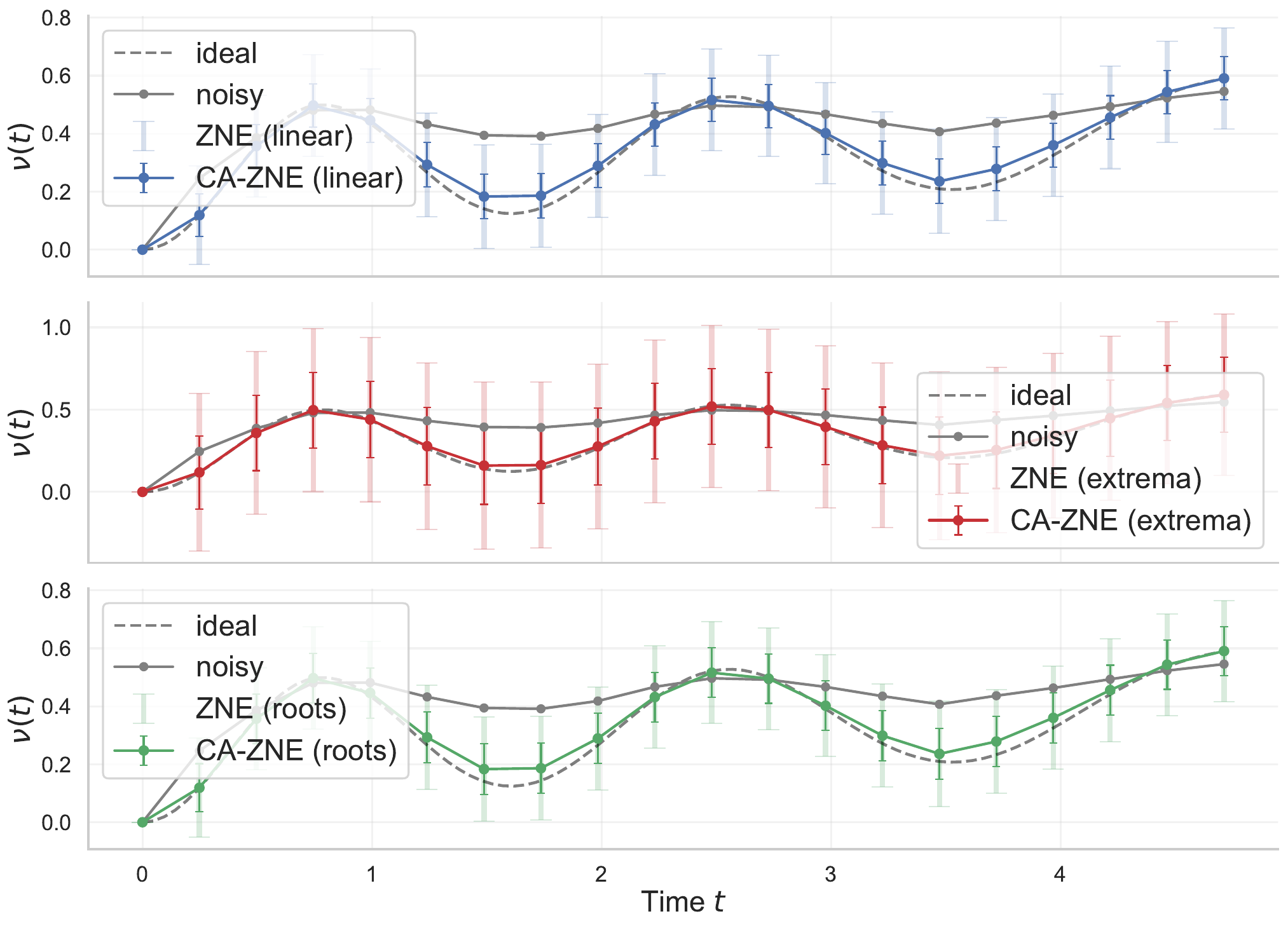}
    \caption{Time dynamics of the particle number of the Schwinger model on 8 qubits. All circuits have been performed for a total of 5 trotter steps and 5 nodes for ZNE. At the noise strengths considered in the simulation ZNE remains slightly biased for the considered number of nodes. The error bars show the standard deviation of the estimate for a shot budget of $N_{\text{shots}} = 10^5$ per time point, which each contain $5$ extrapolation nodes. For all considered spacings CA-ZNE can significantly reduce the error bars.}
    \label{fig:SchwingerModel}
\end{figure}

For quantum simulation, the model can be mapped onto a spin Hamiltonian with long-range couplings~\cite{martinez2016real,ferguson2021measurement}
\begin{equation}
\begin{split}
H = \frac{J}{2} \sum_{n=1}^{S-2} \sum_{k=n+1}^{S-1} (S-k) Z_n Z_k
- \frac{J}{2} \sum_{n=1}^{S-1} (n \bmod 2) \sum_{k=1}^n Z_k \\
+ \frac{\mu}{2} \sum_{n=1}^S (-1)^n Z_n
+ w \sum_{n=1}^{S-1} \left( X_n X_{n+1} + Y_n Y_{n+1} \right).
\end{split}
\end{equation}
Here, $S$ denotes the number of fermionic modes, $\mu$ is the fermion mass, $w = \frac{1}{2a}$, and $J = \frac{g^2 a}{2}$. Both coupling parameters $w$ and $J$ depend on model parameters, namely the lattice spacing $a$ and the coupling strength $g$ of the model.\\
As in the transverse-field Ising case, we study the real-time dynamics of the system using trotterized time evolution. Due to the denser structure of the Hamiltonian of the Schwinger model the circuit obtained by a trotterized time evolution is comparably deep which leads to a rapid spread of entanglement and thus makes it highly susceptible to noise. We construct the time evolution circuits on 8 qubits with 5 trotter steps. The free parameters of our model are chosen to $J = 1.2$, $\mu=0.5$ and $w=1.1$ and the time evolution is simulated from $t=0$ to $t = 3 \pi/2$.\\
To benchmark our approach, we investigate the particle-number density
\begin{equation}
    \nu = \frac{1}{2S}\sum_{n=1}^S \left( (-1)^n Z_n + 1 \right)
\end{equation}
and investigate the time dynamics. The results are presented in Fig~\ref{fig:SchwingerModel}. We investigate ZNE for all three spacings considered in this work. All spacings have been mapped to the interval $[1,3]$ and a total of $5$ nodes was used for ZNE. In contrast for all CA-ZNE experiments we assumed that the first two nodes have been calculated on the quantum computer whereas the last 3 nodes are zero variance estimates. The noise model is given by a local depolarizing layer with $\lambda_0 = 0.0005$ which is applied after each layer of in parallel executable entangling gates.

We find that for all considered spacings ZNE can greatly reduce the bias of the final estimation. The tilted Chebyshev spacing of Ref.~\cite{krebsbach2022optimization} seems to lead to the smallest bias of the three considered spacings, whereas the other two perform comparably but lead to lower variances. Considering CA-ZNE, for all estimates a meaningful reduction in the variance can be achieved generally improving the results.

\subsection{Scaling}
\begin{figure*}[htp]
    \centering
    \includegraphics[width=\linewidth]{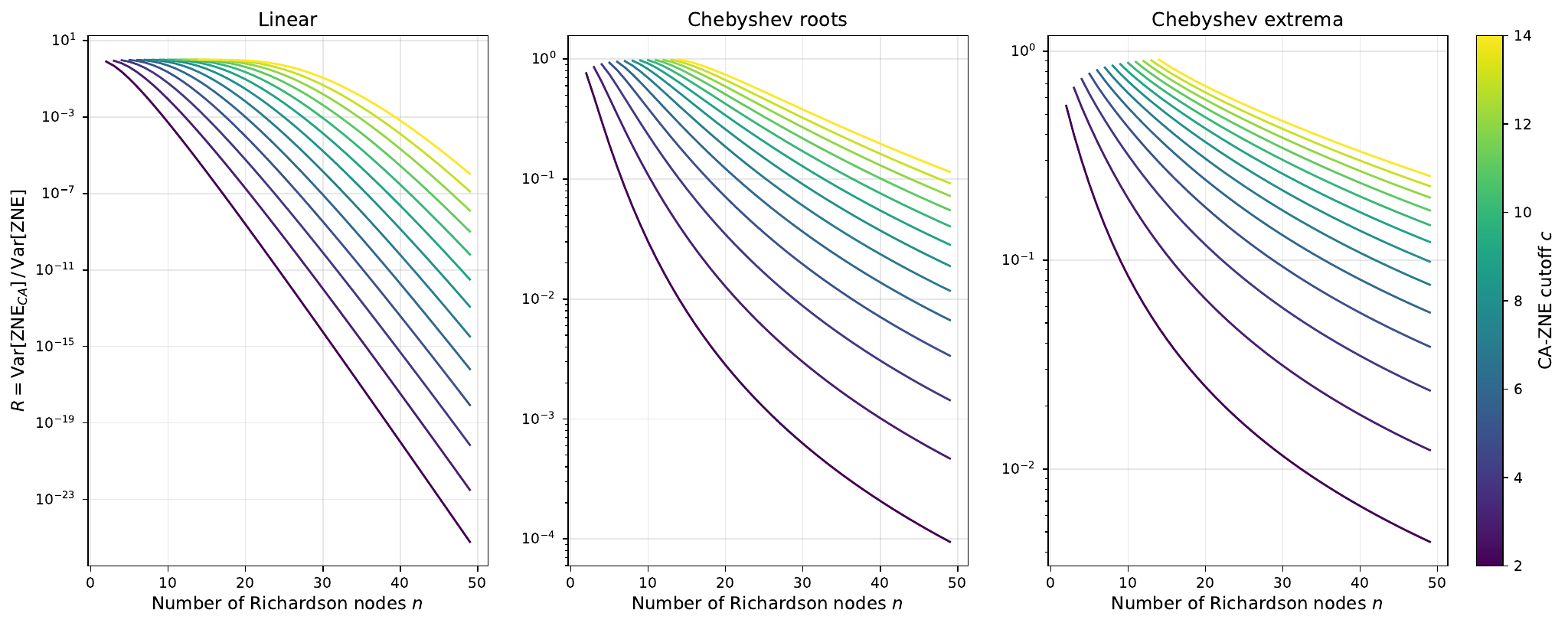}
    \caption{Variance-reduction factor $R=\mathrm{Var}[\hat f_{\mathrm{CA\text{-}ZNE}}]/\mathrm{Var}[\hat f_{\mathrm{ZNE}}]$ as a function of the number of Richardson nodes $n$ for the three node spacings considered in this work, with amplification interval fixed to $[1,B]$ and $B=5$. Curves correspond to different CA-ZNE cutoffs $c$, where nodes with index $i \ge c$ are replaced by classically supplied estimates and the full shot budget is reallocated optimally over the remaining quantum nodes. Smaller values of $R$ indicate stronger variance reduction. For linear spacing, the reduction becomes asymptotically exponential in $n$ at fixed cutoff, whereas for Chebyshev roots and extrema the reduction is substantially more modest.}
    \label{fig:Reduction}
\end{figure*}

\begin{figure*}[htp]
    \centering
    \includegraphics[width=\linewidth]{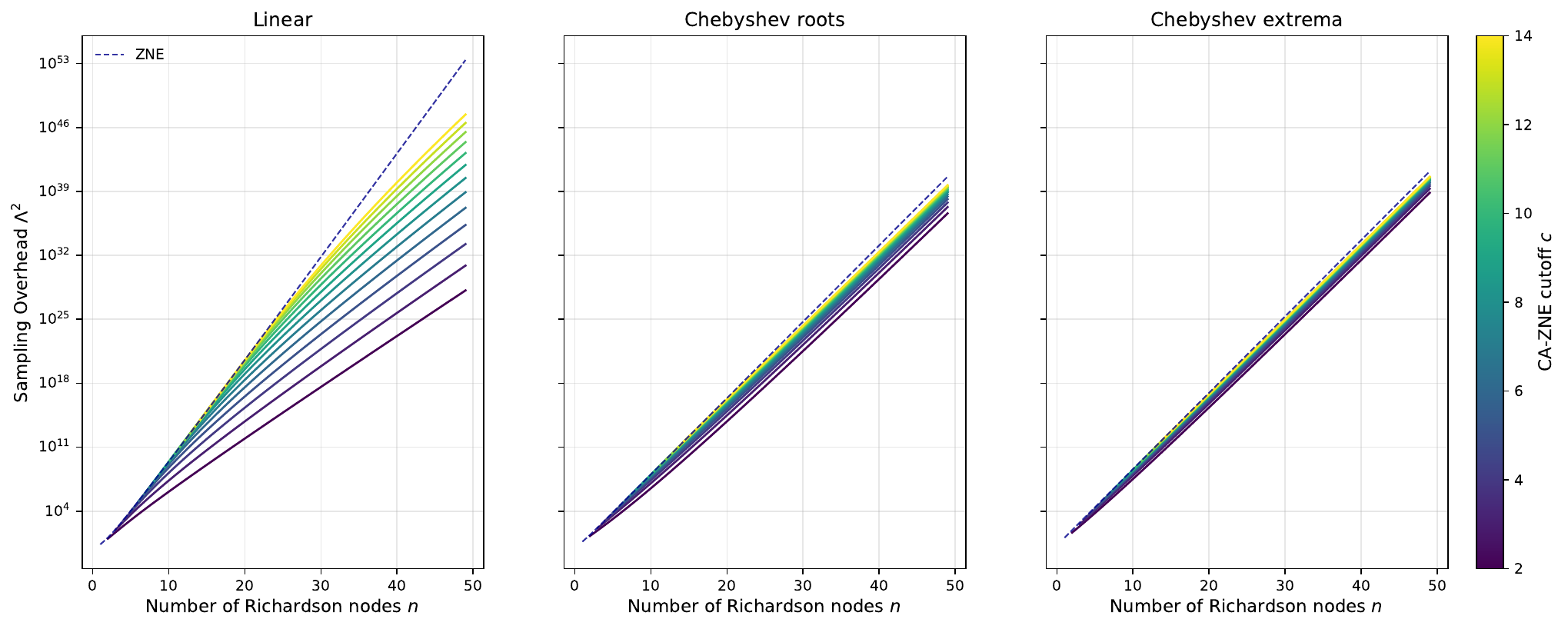}
    \caption{Sampling overhead $\Lambda^2=\left(\sum_i |\gamma_i|\right)^2$ versus the number of Richardson nodes $n$ for standard ZNE and CA-ZNE under the three node spacings considered in this work, with amplification interval fixed to $[1,B]$ and $B=5$. Each CA-ZNE curve corresponds to a cutoff $c$, with nodes $i \ge c$ replaced by classically supplied estimates and the total shot budget redistributed optimally over the remaining quantum nodes. For linear spacing, standard ZNE exhibits the most severe overhead growth, while CA-ZNE can strongly suppress this growth. For the Chebyshev-type spacings, the baseline overhead is already smaller and the additional reduction from CA-ZNE is correspondingly less pronounced.}
    \label{fig:SamplingOverhead}
\end{figure*}

Based on the observed accuracy of our method in predicting the reduction in sampling overhead, we next investigate how this reduction scales with increasing Richardson order. In particular, we study the effect of truncating the Richardson extrapolation for the three node spacings introduced in Sec.~\ref{sec:node_spacing}, namely the linear spacing and the two Chebyshev-type spacings on the interval $[1,5]$. Since we consider all spacings here at the circuit level and do not make any assumptions on the input state $\rho$ or the observable $O$ all results here are for the case $\sigma_{i,O} =1$. Hence for specific input states and observables these results might slightly deviate in practice.\\

The achieved reduction factor $R$ for different extrapolation orders is shown in Fig.~\ref{fig:Reduction}, while the corresponding sampling overheads are reported in Fig.~\ref{fig:SamplingOverhead}. For the linear spacing, given a fixed cutoff $c$, we observe an exponential decrease of the reduction factor $R$ that is essentially independent of the precise value of $c$. We show this explicitly in Appendix~\ref{Appendix:Spacings} for the unbound case. Our findings empirically validate that this carries over to the bounded case. Since our analysis is circuit agnostic, the cutoff is the only quantity that depends on the specific circuit and noise level. This suggests that large sampling variance reductions can be achieved for a broad class of circuits when using linear node spacing. However, if the classical simulation is not exact enough this might affect the bias.

This exponential reduction makes Richardson extrapolation with linear node spacing competitive in terms of the sampling overhead, despite its otherwise prohibitive scaling (see Fig.~\ref{fig:SamplingOverhead}).

For the optimal Chebyshev spacing introduced in Ref.~\cite{krebsbach2022optimization}, the reduction is more modest but still meaningful, reaching a factor of approximately $10^{-3}$ at larger extrapolation orders $n$. The corresponding sampling overheads are shown in Fig.~\ref{fig:SamplingOverhead}. This behavior can be explained by the distribution of the interpolation weights $|\gamma_i|$, which place greater emphasis on nodes $x_i$ at the low-noise end of the spacing. Under importance sampling, most of the shot budget is therefore assigned to these nodes, while the high-noise nodes contribute comparatively little. We illustrate this graphically for all spacings in App.~\ref{Appendix:Spacings}.

The Chebyshev root spacing used in Ref.~\cite{mohammadipour2025direct} yields more favorable reductions than the Chebyshev spacing of Ref.~\cite{krebsbach2022optimization}. Although it still does not lead to an exponential reduction, it lowers the variance by one to four orders of magnitude, making CA-ZNE a promising approach for this choice of spacing.

Overall, our results suggest that, when augmenting Richardson extrapolation with classical nodes, spacings that place greater emphasis on high-noise nodes may be more advantageous for reducing sampling overhead than spacings concentrated on low-noise nodes.

\section{Discussion}
\label{sec:Discussion}
During the experiments we have relied on simulating some of the nodes classically.
The method presented in this work thus raises a fundamental question: 
\textit{What is the value of the quantum computer in zero-noise 
extrapolation?} If we can simulate many nodes classically, why should we include the quantum nodes to begin with?\\
Corollary~1 of Ref.~\cite{schuster2025polynomial} 
establishes that any error mitigation strategy that succeeds 
in polynomial time on most input states can be replaced by a 
classical simulation of equivalent accuracy. Intuitively, this 
is because observables whose ideal expectation values are 
recoverable via error mitigation must be dominated by low-weight 
Pauli operators, 
which are simultaneously easy to simulate classically. Conversely, 
for observables with significant high-weight Pauli contributions, 
error mitigation fundamentally cannot recover the ideal 
expectation value, since these contributions are exponentially 
suppressed by noise.

Our method targets the intermediate regime of observables with 
both low-weight and high-weight Pauli contributions. At high 
noise levels, the high-weight contributions are exponentially 
damped, and classical simulation provides accurate, zero-variance 
estimates. At low noise levels, however, high-weight Pauli 
operators remain partially visible, damped by $e^{-\gamma w}$ 
with small $\gamma$, and contribute meaningfully to the 
expectation value. Simulating these contributions classically 
would require larger truncation thresholds, yielding
run-times that are impractical despite being formally 
quasi-polynomial~\cite{schuster2025polynomial}.\\

In contrast, the 
quantum computer measures the full expectation value directly, 
including all high-weight contributions, at a cost determined 
solely by the shot budget. If the noise level is sufficiently low the quantum computer is able to extract meaningful signal that remains practically out of reach for the classical simulation.\\
This separation is especially promising in combination with 
partial probabilistic error cancellation~\cite{cai2021multi, 
mari2021extending}. By redistributing the shot budget freed 
from the classically simulated high-noise nodes, one can apply 
partial PEC at the quantum nodes. This could extend the 
accessible weight range beyond what standard noisy measurements 
provide, amplifying the quantum advantage of the hybrid scheme.\\

Another important caveat is that including an increasingly large amount of classical nodes to reduce the bias does not guarantee continued improvement. As recently shown in Ref.~\cite{koester2026benchmarking}, once the amplified expectation values at large noise levels have reached an observable-dependent floor, Richardson extrapolation may no longer recover the underlying physical signal and instead produce artefactual improvements. For CA-ZNE, this means that replacing high-noise quantum nodes by classical estimates can reduce variance, but does not by itself prevent this failure mode. Hence, the nodes in both ZNE and CA-ZNE should be chosen within an interval $[1,B]$
such that the maximum amplified noise level $\lambda_0B$ remains in a signal-retaining regime, i.e., before the expectation values collapse to the noise floor.

\section{Conclusion \& Outlook}
\label{sec:CO}
In this work we introduced and investigated Classically Augmented
Zero-Noise Extrapolation (CA-ZNE). Our results show that replacing
high-noise Richardson nodes with classically simulated anchor points
can substantially reduce the sampling variance of the extrapolation.
This reduction can be exponential for linear node spacings and more modest for others.

While these reductions are striking in principle, they must be
interpreted with care. The method does not remove error but reallocates it, trading sampling variance for classical simulation bias. Because both the variance and the bias are governed by the interpolation weights $|\gamma_i|$, the classical estimates must become increasingly accurate as more nodes, or equivalently more cumulative interpolation weight, are offloaded to classical simulation. The practical value of CA-ZNE therefore hinges on whether the classical bias can be kept below the variance it eliminates.

Our analysis of node spacings further suggests that the spacings known to be optimal for standard ZNE need not remain optimal once
zero-sampling-variance anchor points are available. Deriving optimal
spacings for this augmented setting is a natural and worthwhile
direction for future work. Equally important is a certifiable,
data-driven cutoff-selection protocol that guarantees the classical
augmentation does not increase the overall error.

\bibliographystyle{apsrev4-1} 
\bibliography{references} 

\section*{Acknowledgements}
This work was supported by the research project “Zentrum für Angewandtes Quantencomputing” (ZAQC), funded by the Hessian Ministry for Digital Strategy and Innovation and the Hessian Ministry of Higher Education, Research and the Arts.\\
The author would like to thank Matthias Heller and Paul Haubenwallner for useful discussions and for suggesting the Schwinger model as a test bed.

\section*{Author Contributions}
The author conceived the study, developed the methodology, derived the
analytical results, implemented and validated the numerical methods,
performed the simulations, analyzed the results, prepared the figures,
and wrote and revised the manuscript.

Generative-AI tools (Claude Opus 5 and GPT-5.6 Sol) were used for
language editing, identifying passages and technical points requiring
additional verification, and generating prototype code snippets. The
author reviewed and tested all code used to produce the reported
results, verified the derivations and numerical results and takes full responsibility
for the content of the manuscript.

\section*{Data availability}
All data in this manuscript is available from the author upon request.

\appendix
\appendix

\section{Upper bound of the SPL model}
\label{Appendix:NoiseModel}
In this appendix, we show that the damping induced by the SPL model on an observable $O$ can be upper-bounded, in Frobenius norm, by that of a local depolarizing channel $\mathcal{D}=\bigotimes_{i=1}^n \mathcal{D}_i$
\begin{equation}
\label{eq:Depol}
    \|\mathcal{N}(O)\|_F^2 \le \|\mathcal{D}(O)\|_F^2,
\end{equation}
with effective damping rate $\nu=4\lambda_{\min}$.\\

\paragraph*{Proof.}
 Without loss of generality we consider Pauli operators forming an orthonormal basis with respect to the Hilbert-Schmidt (Frobenius)-norm $\text{Tr}(P_i^\dagger  P_j) = \delta_{ij}$. Hence any observable $O=\sum_i c_i P_i$ has
 \begin{equation}
 ||O||_F^2=\sum_i |c_i|^2.
 \end{equation}
We consider a sparse Pauli-Lindblad noise model~\cite{van2023probabilistic} to capture the noise of a quantum device. The SPL model is generated by a Lindbladian of the form
\begin{equation}
    \mathcal{L}(\rho) =\sum_{k\in \mathcal{K}} \lambda_k (P_k\rho P_k - \rho)
\end{equation}
for some rates $\lambda_k$. The set of generators $\mathcal{K}$ contains all errors that are considered in the noise model (usually 1- and 2-local Pauli generators). We assume that under realistic conditions all single qubit Pauli operators are contained in the model and all error rates for these terms are non-zero $\lambda_k >0$.
The relation between the rates $\lambda_k$ and the weights $w_k$ in Eq.~\eqref{eq:SPL_noise} is given by $w_k = (1+e^{-2\lambda_k})/2$.
The effect of the noise channel $ \mathcal{N}(\rho) = e^{\mathcal{L}(\rho)}$, generated by the Lindbladian $\mathcal{L}$, on a Pauli $P$ leads to an exponential damping of its scalar coefficient
\begin{equation}
\label{eq:damp}
     \mathcal{N}(P) = e^{-2\sum_{k\in \mathcal{K}} \lambda_k \langle P, P_k\rangle}P
\end{equation}
where $\langle P, P_k\rangle$ denotes the symplectic inner product which is $0$ whenever $[P,P_k] = 0$ and $1$ otherwise~\cite{van2023probabilistic}.\\
For a Pauli observable we obviously have $||\mathcal{N}(P)||_F^2 \leq ||P||_F^2$. It hence follows for the considered observable $O$ that
\begin{equation}
    || \mathcal{N}(O)||_F^2 = \sum_i |c_i|^2e^{-4\sum_{k\in \mathcal{K}} \lambda_k \langle P_i, P_k\rangle} \leq ||O||_F^2 
\end{equation}
since the Paulis are orthogonal under the Hilbert-Schmidt inner product $\text{Tr}(P_j^\dagger P_k) = \delta_{jk}$.
We can upper bound the damping of an observable in the Frobenius norm by choosing the set $\mathcal{K}_\text{local} $ containing only the 1-local Pauli terms resulting in the local SPL channel $ \mathcal{N}_\text{local}$. Since we only remove positive coefficients from the sum in Eq.~\eqref{eq:damp} it follows directly that 
\begin{equation}
    || \mathcal{N}(P)||_F^2 \leq || \mathcal{N}_\text{local}(P)||_F^2
\end{equation}
since
\begin{equation}
e^{-2\sum_{k\in \mathcal{K}}\lambda_k \langle P,P_k\rangle}
\le
e^{-2\sum_{k\in \mathcal{K}_{\mathrm{local}}} \lambda_k \langle P,P_k\rangle}.
\end{equation}
Where the equality holds only if the noise model was 1-local from the start. Further, the generators can be grouped into $n$ sets of the 3 single qubit Pauli operators, where each of these sets of super operators are mutually commuting. Hence the SPL model can be expressed as a tensor product of local channels
\begin{equation}
     \mathcal{N}_\text{local}= \bigotimes_{i=1}^n \mathcal{N}_i.
\end{equation}
Again we can upper bound the Frobenius norm after damping by setting all coefficients $\lambda_k$ to the smallest coefficient $\lambda_\text{min} =\underset{k}{\text{min}}~ \{\lambda_k > 0\} $ since
\begin{equation}
e^{-2\sum_{k\in \mathcal{K}_{\mathrm{local}}}\lambda_k \langle P,P_k\rangle}
\le
e^{-2\sum_{k\in \mathcal{K}_{\mathrm{local}}} \lambda_{\min} \langle P,P_k\rangle},
\end{equation}
leading to the model
\begin{equation}
     \mathcal{N}_{i,\text{min}}(\rho) = \prod_{\mathcal{P}_k \in \{\mathcal{X},\mathcal{Y},\mathcal{Z}\}} (w \mathcal{I}(\cdot) + (1-w) \mathcal{P}_k(\cdot)) \rho
\end{equation}
where the weights $w$ are related to the minimum error rate $\lambda_\text{min}$ via $ w = (1+e^{-2\lambda_\text{min}})/2$.
Finally, by explicitly expanding the product structure one obtains
\begin{equation}
\label{eq:depol1}
    \mathcal{N}_{i,\text{min}}(\rho) =(1-3p)\rho + p(X\rho X + Y\rho Y + Z\rho Z) 
\end{equation}
 which is an explicit representation of the single qubit depolarizing channel $\mathcal{D}_i(\rho)$. A short calculation shows that $p = w-w^2$. It follows that the magnitude of a Pauli term damped by the SPL model is upper bounded by a layer of 1-local depolarizing channels
\begin{equation}
    ||\mathcal{N}(O)||_F^2 \leq ||\mathcal{D}(O)||_F^2
\end{equation}
where $\mathcal{D} =\bigotimes_{i=1}^n \mathcal{D}_i$. The effective damping coefficient is hereby given as $p=\frac{1}{4}(1-e^{-4\lambda_\text{min}})$. Equivalently the depolarizing channel can be written as
\begin{equation}
\label{eq:depol2}
    \mathcal{D}_i(\rho) =e^{-\nu}\rho + (1-e^{-\nu}) (\text{Tr}_i(\rho)\otimes\frac{\mathbb{I}}{2}).
\end{equation}
where $\text{Tr}_i$ denotes the partial trace over qubit $i$ and $\frac{\mathbb{I}}{2}$ denotes the single qubit maximally mixed state. By comparing the eigenvalues of an arbitrary single qubit Pauli it can be seen that
\begin{equation}
    \mathcal{D}_i(P)  = (1-4p) P = e^{-\nu} P
\end{equation}
where we utilized equations~\eqref{eq:depol1} and \eqref{eq:depol2} respectively. We can thus identify
\begin{equation}
    \nu = 4\lambda_\text{min}
\end{equation}
which proves proposition~\eqref{eq:Depol}.

\section{Bounds on the Bias}
\label{Appendix:Bias}
\paragraph*{Proof.} We derive an upper bound on the expected bias introduced when a subset of Richardson nodes is replaced by classically simulated estimates. The expected bias of our methods is given by
\begin{equation}
    \begin{split}
    \mathbb{E}[\hat{f}_\text{aug}(0)] - \langle O\rangle_0
    &=  \left( \sum_{i=0}^{n-1} \langle O\rangle_i \gamma_i -  \langle O\rangle_0\right)\\
    &+\left( \sum_{i=0}^{c-1} (\mathbb{E}[\hat{O}_{i}^\text{quant}] -  \langle O\rangle_i )\gamma_i\right )\\
    &+ \left( \sum_{i=c}^{n-1}  (\mathbb{E}[\hat{O}_{i}^\text{class}] -  \langle O\rangle_i )\gamma_i\right)\\
    \end{split}
\end{equation}
where we have added the identity $\sum_{i=0}^{n-1} \langle O\rangle_i \gamma_i - \sum_{i=0}^{n-1} \langle O\rangle_i \gamma_i$. The bias hence decomposes into three distinct terms, the extrapolation error $\Delta_\text{extrap} = \sum_{i=0}^{n-1} \langle O\rangle_i \gamma_i -  \langle O\rangle_0$, the sampling error of the quantum computer $\Delta_\text{sampl} = \sum_{i=0}^{c-1} (\hat{O}_{i}^\text{quant} -  \langle O\rangle_i )\gamma_i$ and the bias introduced by the classical nodes $\Delta_\text{approx} = \sum_{i=c}^{n-1} (\hat{O}_{i}^\text{class} -  \langle O\rangle_i )\gamma_i$.\\
Since we consider the expectation of the bias $\mathbb{E}[f_\text{aug}(0)] - \langle O\rangle_0$ and the quantum nodes are expected to be unbiased we can assume that $\mathbb{E}\left[ \sum_{i=0}^{c-1} (\hat{O}_{i}^\text{quant} -  \langle O\rangle_i )\gamma_i\right ] = 0$. The bias is thus only given by the extrapolation bias as well as the classical simulation bias. The extrapolation bias has already been investigated in several works~\cite{krebsbach2022optimization, mohammadipour2025direct}. Following the theory on polynomial interpolation the extrapolation bias for some $\xi \in [0, x_{n-1}]$ is given by
\begin{equation}
    \left|f(0) - \langle O\rangle_0 \right| \leq \left|\frac{ g^{(n)}(\xi)}{(n)!}\prod_{j=0}^{n-1} x_j \right|
\end{equation}
where we assume that the true noisy expectation value function $g(x) = \langle O \rangle_{\lambda_0 x}$ is $n$ times differentiable.
The extrapolation bias hence directly relies on the choice of nodes $x_j$.

To bound the bias introduced by the simulation we consider
\begin{equation}
     \left|\sum_{i=c}^{n-1} (\hat{O}_{i}^\text{class} -   \langle O\rangle_i )\gamma_i \right|\leq \sum_{i=c}^{n-1}  |\delta_i|~ |\gamma_i|
\end{equation}
where $\delta_i$ is the truncation bias of the classical simulation. By utilizing the algorithm of~\cite{schuster2025polynomial} this can be bounded by
\begin{equation}
    |\delta_i| \leq e^{-\nu_i(l+1)} \sqrt{d+1}||O||_F
\end{equation}
where $\nu_i=\nu_0 x_i$ is the effective noise rate of node $i$.
Using these definitions the full bias of our estimate is bounded by
\begin{equation}
\begin{split}
    \left|\mathbb{E}[\hat{f}_\text{aug}(0)] - \langle O\rangle_0 \right| &\leq
     \left|\frac{ g^{(n)}(\xi)}{(n)!}\prod_{j=0}^{n-1} x_j \right| \\
    &+\sqrt{d+1}||O||_F\sum_{i=c}^{n-1} |e^{-\nu_i(l+1)}|~|  \gamma_i|
\end{split}
\end{equation}
where in the last equation we have made implicit use of the triangle inequality.

\section{Interpolation weight distribution}
\label{Appendix:Spacings}

\begin{figure*}[htp]
    \centering
    \includegraphics[width=\linewidth]{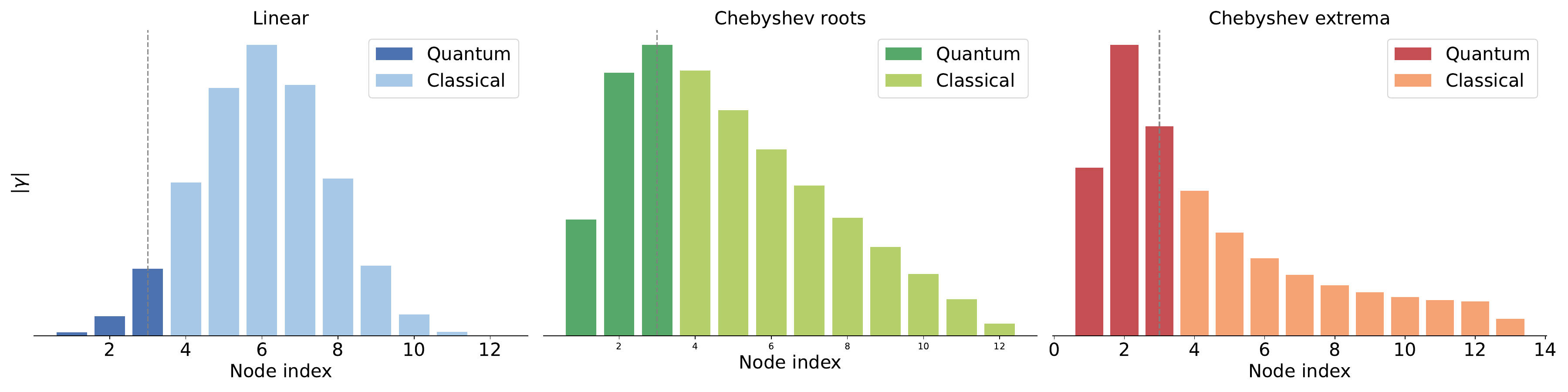}
    \caption{Distribution of the interpolation weight $\gamma$ for the three considered spacings. The data shows 10 nodes with a cutoff for the quantum computer of $c=3$ at $B = 10$. As can be seen, the linear spacing puts more emphasis on nodes in the middle of the spacing, whereas the Chebyshev spacings put more emphasis on the low noise nodes.}
    \label{fig:Gamma_dist}
\end{figure*}

\begin{figure*}[htp]
    \centering
    \includegraphics[width=\linewidth]{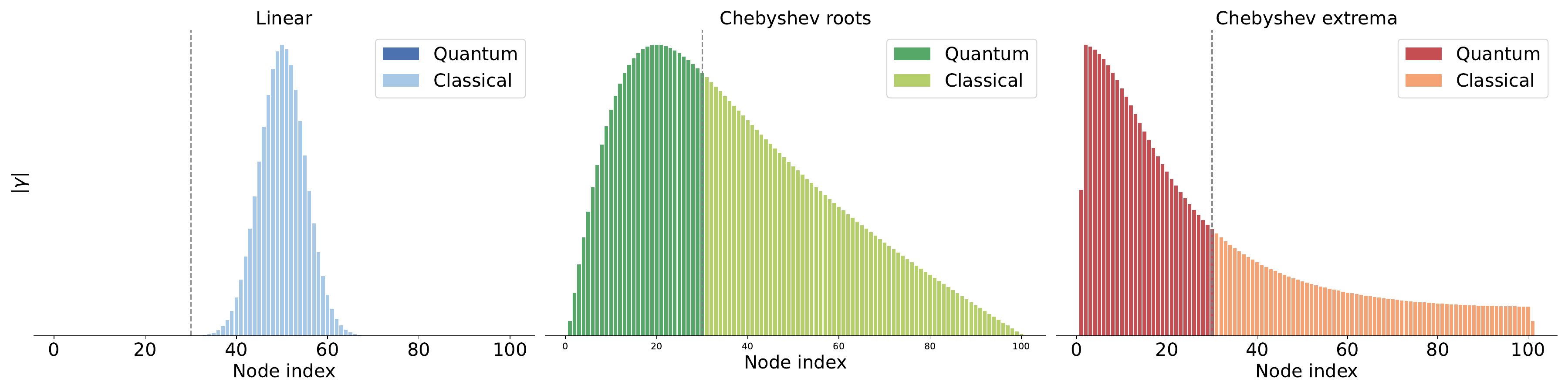}
    \caption{Same data as in Fig.~\ref{fig:Gamma_dist} for a higher number of nodes $n=100$ at $B = 10$. As highlighted by the data the distribution of weights approaches different distributions with different means. For the selected cutoff of $c=30$ it can be seen that most (linear), almost none (Chebyshev extrema) or some (Chebyshev roots) of the large-weight nodes can be outsourced to the classical computer.}
    \label{fig:Gamma_dist_lim}
\end{figure*}

In this appendix, we briefly explain why CA-ZNE yields an exponential reduction in sampling overhead for linearly spaced nodes when the cutoff is held fixed. We provide this analysis for the direct, unbounded spacing for simplicity. Similar results can be obtained for the bounded spacing~\cite{mohammadipour2025direct} in the range $[1,B]$.
For the interpolation weight at node $i$ is generally given by 
\begin{equation}
    \gamma_j =\prod_{k\neq j}^{n-1} \frac{x_k}{x_k -x_j}
\end{equation}
when considering a linear spacing $x_j = 1 + j$ this reduces to
\begin{equation}
    \gamma_j =\prod_{k\neq j}^{n-1} \frac{1+k}{k-j}.
\end{equation}
This can be further simplified to
\begin{equation}
     \gamma_j =  (-1)^j \frac{(n)!}{(j+1)!((n-1)-j)!}.
\end{equation}
This result can be seen as the special case of $h = 1$ already obtained in ref.~\cite{mohammadipour2025direct}. By identifying the binomial coefficient it directly follows that the weights for the linear spacing are binomially distributed
\begin{equation}
    |\gamma_j| =\binom{n}{j+1}.
\end{equation}
Using this definition the sampling overhead is given by
\begin{equation}
\label{eq:binom}
    \Lambda = \sum_{j=0}^{n-1} \binom{n}{j+1} = \sum_{m=1}^{n} \binom{n}{m}
\end{equation}
The value of the series defined by Eq.~\eqref{eq:binom} is well known and follows from the binomial theorem
\begin{equation}
    \Lambda = 2^{n} - 1.
\end{equation}
This explains the exponential sampling overhead that is incurred by the linear spacing.\\

We can now shift our attention to the improvement obtained by replacing the last $n-c$ nodes with classical zero variance nodes.
Following Eq.~\eqref{eq:red} the reduction is given by
\begin{equation}
\label{eq:exp_red}
    R = \left(\frac{\sum_{m=1}^{c} \binom{n}{m}}{2^{n} - 1} \right)^2
\end{equation}
for a fixed cutoff $c$. To show that the reduction is exponential it suffices to show that the numerator of Eq.~\eqref{eq:exp_red} is at most polynomial in $n$. For $c < n/2$ we can assume that
\begin{equation}
    \sum_{m=1}^{c} \binom{n}{m} = \binom{n}{c} (1+O(1/n)).
\end{equation}
Thus the dominant polynomial contribution is given by the last summand
\begin{equation}
    \binom{n}{c} = \frac{(n)!}{(c)!(n-c)!} = O(n^{c})
\end{equation}
which is a polynomial of degree $c$.
This explains that the achieved reduction is~\textit{asymptotically} exponential.\\

We show a modest reduction for the Chebyshev extrema and a reasonable reduction for the Chebyshev roots. Since there is no closed form for the interpolation weights, as for the linear spacing, we give a graphical representation of the spacing of interpolation weights for all considered spacings in figs.~\ref{fig:Gamma_dist} and~\ref{fig:Gamma_dist_lim}

\end{document}